\documentclass[%
 reprint,
superscriptaddress,
nofootinbib,
 amsmath,amssymb,
 aps,
 prl,
]{revtex4-2}

\usepackage{graphicx}
\usepackage{dcolumn}
\usepackage{bm}

\usepackage{amsmath,amssymb,bm}
\usepackage{mathrsfs}
\usepackage{textcomp}
\usepackage{commath}
\usepackage{mathtools}
\usepackage{natbib}
\usepackage{placeins}
\usepackage{graphicx} 
\graphicspath{{figures/}}  
\usepackage[usenames]{color} 

\newcommand{\Ca}{\mathrm{Ca}}

\newcommand{\evn}{\mathrm{E_V}}
\newcommand{\evnc}{\mathrm{E^C_V}}

\usepackage[normalem]{ulem}

\begin{document}

\preprint{APS/123-QED}

\title{Flapping instability of elastic disks in Stokes flows.}

\author{Yijiang Yu}\email{Equal contribution}
\affiliation{
Department of Chemical and Biological Engineering\\
University of Wisconsin-Madison, Madison, WI 53706-1691
}
\author{Hugo Perrin}\email{Equal contribution}
\affiliation{
Process and Energy Department, 3ME Faculty of Mechanical, Maritime and Materials Engineering,
TU Delft, 2628 CD Delft, The Netherlands
}
\affiliation{
Paris-Saclay University, FAST, 91405, Orsay, France
}
\author{Michael D. Graham}\email{Corresponding author. E-mail: mdgraham@wisc.edu}
\affiliation{
Department of Chemical and Biological Engineering\\
University of Wisconsin-Madison, Madison, WI 53706-1691
}

\author{Lorenzo Botto}\email{Corresponding author. E-mail: l.botto@tudelft.nl}
\affiliation{
Process and Energy Department, 3ME Faculty of Mechanical, Maritime and Materials Engineering,
TU Delft, 2628 CD Delft, The Netherlands
}


\date{\today}

\begin{abstract}
 Fluid-structure interactions at low Reynolds number can lead to a much richer phenomenology than previously expected.  Here, we study the dynamics of a freely suspended, thin elastic disk in a shear flow, where the plane of the disk is initially \emph{parallel} to the flow plane. Using a combination of experiments and simulations, we demonstrate that beyond a critical flow strength the disk deforms, performing flapping dynamics, in which the disk curves up and down periodically relative to the horizontal shear plane. The bifurcation diagram obtained by simulation reveals several oscillatory solutions, including a wiggling motion that is predicted by a linear stability analysis. The flapping dynamics is shown to be a subcritical instability whose key ingredient is the finite extensibility of the disk. The behavior we observe has implications for emerging investigations on the flow dynamics of sheet-like particles, such as 2D polymers and 2D crystalline materials immersed in viscous fluids.


\end{abstract}

\maketitle


A cardinal problem in hydrodynamics is the prediction of the dynamics of small anisotropic objects suspended in a shear flow. Besides having contributed to the development of early theories on the atomistic nature of matter \cite{FPerrin36,Burgers1995}, this problem has implications that span practically all areas of physics and biology \cite{du2019dynamics}. The simplest setting in which such dynamics can be investigated is the low-Reynolds-number motion of an inertialess non-Brownian particle. More than one century ago, a foundational theory formulated by Jeffery \cite{jeffery1922motion} predicted that a rigid spheroid in a simple shear flow undergoes periodic orbits with dynamics only determined by the initial orientation.  In the absence of Brownian motion the orientational trajectories maintain the basic features of the underlying governing equations, such as linearity and time-reversibility. However, it is now appreciated that  when the object is deformable the coupling between flow and deformability can break such symmetry properties, leading to loss of time reversibility \cite{PRLCantat99,Lorz2000WeaklyAV, Bureau2023aa,PRL2002Viallat,PRF25Stone}, non-linear dependence on the controlling parameters \cite{ARFLindner,lindstrom2007simulation, harasim2013direct, du2019dynamics, slowicka2020flexible,PRK2006Ripoll} and oscillating dynamics \cite{kantsler2005orientation, misbah2006vacillating, bagchi2009dynamics, biben2011three,abkarian2007swinging, dupire2012full, sinha2015dynamics}. This fluid-structure interactions problem has been studied extensively for flexible fibers (synthetic polymers, DNA, protein nanofibrils \cite{santos2023flow}, etc.) and anisotropic vesicles and cells of small aspect ratio \cite{abkarian2007swinging,Lorz2000WeaklyAV,kantsler2005orientation}.  However, there is a need to understand of how flexible sheet-like particles of large aspect ratio orient and deform in shear flow \cite{verhille2022deformability, botto2019toward,islam2022exfoliation, reddy2018rheo}. This need is motivated by the enormous growth of interest in 2D crystalline  nanomaterials   (e.g. graphene, InSe,
WSe2, BN, etc.) and 2D polymers, and their processing in liquids. 

Deformable thin sheets oriented with their normal in the shear flow plane exhibit periodic buckling deformations similar to those exhibited by fibers \cite{ARFLindner} beyond a buckling threshold \cite{LINGARD1974119,perrin2023hydrodynamic,silmore2021buckling,funkenbusch2024dynamics,funkenbusch2024shear}. The competition between elastic and viscous forces is quantified by the elasto-viscous number $\evn = \eta \dot \gamma a^3/K_b$, where $\eta $ is the viscosity of the fluid, $\dot \gamma$ is the imposed shear rate, $a$ is the half-length of the sheet and $K_b$ is the bending stiffness. These periodic deformations are not surprising: they can be explained by the periodic alignment of the sheet along angular regions of the shear flow where the hydrodynamic forces are alternatively compressive and extensional \cite{du2019dynamics}. However, the orientation in which the plane of the sheet is parallel to the plane of the flow is more subtle. The usual assumption is that for this configuration symmetry arguments yield a trivial rolling motion, with no particularly interesting dynamical features. However, this argument overlooks the possible emergence of elasto-viscous instabilities (even under simple compressive loads, plates display a family of symmetry-breaking buckling bifurcation~\cite{timoshenko1959theory,audoly2000elasticity}). The fundamental question is whether elasto-viscous instabilities lead to a change in average orientation of the sheet or to the convergence towards a new dynamical state in which deformation plays a primary role. 

\begin{figure}[h]
\centering
\includegraphics[width=0.5\textwidth]{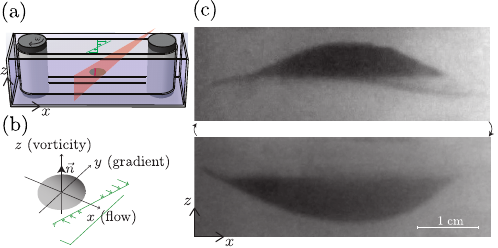}
\caption{(a) Experimental setup. The red triangular region represents the inclined laser sheet.  The intersection of the laser sheet with the disk enables us to measure the local height and out-of-plane deformation of the sheet. (b) Coordinate system. A disk is initially placed inside the plane $z = 0$ with its normal $\overrightarrow{n}$ along $z$. (c) Experimental side images of a flapping disk at two selected instants.}
\label{fig:setup}
\end{figure}

In this letter, we report experimental and numerical observations of a thin elastic disk suspended in a shear flow and initially oriented parallel to the flow plane. We demonstrate that if $\evn$ is small, the disk remains flat and rotates according to the predictions of Jeffery's theory for rigid objects. However, beyond a threshold value of $\evn$, a regime emerges whereby the disk curves periodically while rotating (Fig.\ref{fig:setup}(c)). We called this the \emph{flapping} regime. A numerical linear stability analysis enables us to precisely determine the thresholds of the instabilities. The numerical results reveal an unexpectedly complex bifurcation diagram as a function of $\evn$. For example, numerics show a stable branch of the bifurcation diagram corresponding to an asymmetric \emph{wiggling} motion regime which is not observed in experiments. Surprisingly, the periodic wiggling and flapping deformations do not lead to a reorientation of the disk, and are sustained for a long time. For large values of $\evn$ a reorientation is observed.

These results contribute to an improved understanding of the dynamics of thin structures in flow. Recent work has explored the dynamics of thin sheets in shear flow~\cite{xu2014brownian, dutta2017dynamics, silmore2021buckling, silmore2022thermally, kamal2020hydrodynamic, salussolia2022simulation, leong2010stability, perrin2023hydrodynamic,kamal2021effect,qi2025unraveling} by addressing the importance of initial orientation, surface slip, and deformability. In extensional flows, highly stretchable sheets show a coil-stretch-like transition with bistable conformations, while sufficiently flexible sheets in uniaxial extensional flow exhibit a complex wrinkling instability due to compressive hydrodynamic stresses ~\cite{yu2021coil,yu2022wrinkling}. Non-Brownian flexible sheets can reach a global steady state in shear flow where the sheet rests within the flow-vorticity plane, as a flow-aligned state~\cite{silmore2021buckling}. Deformable graphene sheets, which display substantial hydrodynamic slip, have been shown to reach an alignment with the normal in the flow plane regardless of the initial orientation, and in this orientation reach a stable alignment rather than rotating \cite{agrawal2025negative,gravelle2021violations,gravelle2025effect}.  However, so far no oscillatory dynamics have been observed for no-slip elastic sheets for flow parallel orientations.

{Our experiments are carried out with circular disks cut from thin films of silicon elastomer, with disk radius $a \simeq 1.14\;\rm{cm}$ and thickness $h\simeq 180\;\rm{\mu m}$. The bending stiffness of the sheet is calculated as $K_b = E h^3 / 12 (1-\nu^2)$, where $\nu\simeq 0.5$ is the Poisson ratio and the 3D Young modulus $E\simeq 370 \;\rm{kPa}$ is measured with a rheometer (Anton Paar MCR 501), see Supplemental Materials. The shear cell (Fig.\ref{fig:setup}a) is composed of a belt driven by two co-rotating cylinders of diameter $6\; \rm{cm}$. A  motor, connected to one of the cylinders, imposes a controlled shear rate in the range $\dot \gamma =  0.4- 10\;  \rm{s^{-1}}$. The sheets were placed in glycerol (viscosity $\eta \simeq 1\;\rm{Pa.s}$,  density  $\rho\simeq 1.2\times10^{3}\;\rm{kg.m^{-3}}$). To prevent settling, the silicon elastomer is density matched with the fluid by adding carbon black to the silicon prior to the polymerization reaction. The maximum Reynolds number $Re = \rho \dot \gamma a^2/\eta$ is of order 1 at the maximum shear rate. In each experiment, the disk was set with its normal vector along the vorticity direction of the undisturbed flow by  use of tweezers. Optical measurements with a camera were carried out from the top, i.e. along the vorticity direction.  The deformation of the disk are measured from the distortion of a laser line intersecting the disk at an assigned angle. We extracted from the images the instantaneous disk curvature $\kappa(t)$ by fitting a parabola to the laser line -- see Fig.\ref{fig:setup}(b). Measurements were performed with a time resolution of $0.01\;\rm{s}$ and with a spatial resolution of $25\;\rm{\mu m / pixel}$.
}

The simulations are carried out with a regularized Stokeslet method~\cite{cortez2005method, graham2018microhydrodynamics}. The method solves the steady Stokes equation coupled to the elastic deformation of a sheet of zero hydrodynamic thickness, accounting for hydrodynamic interactions between different portions of the sheet. We simulate the sheet by  tracking  a mesh of nodes labeled $i$ at positions $\mathbf{X}_i$ that move as material points on the sheet surface. The elastic responses of the sheet include in-plane stretching and out-of-plane bending~\cite{charrier1989free,fedosov2010multiscale,yu2021coil,yu2022wrinkling}. The in-plane stretching is modeled as a neo-Hookean solid with strain energy density   $W = G(I-3)/2$, where $G$ is the 2D shear modulus and $I$ is the first invariant of the Cauchy deformation tensor. For comparison with the experiments, the 2D shear modulus is related to material properties and thickness according to $G=E h / 2 (1+\nu) $  where $E$ is the 3D Young modulus and $h$ the thickness of the disk. The out-of-plane bending is modeled by a bending energy that is calculated from the out-of-plane deflection of the dihedral angles $\theta_{\alpha\beta}$ between neighboring elements:
\begin{equation}
    E_b = \sum_{\mathrm{adj}\ \alpha,\beta}k_b \left[1-\cos(\theta_{\alpha\beta} )\right].
\label{eq: bending}
\end{equation}
Here, $k_b$ is a bending constant. This parameter was related to the bending stiffness $K_B$ according to $k_b/K_B \simeq 1.09$, a number that was obtained by comparing Eq. \eqref{eq: bending} with the bending energy of a disk bent into a surface of known constant curvature \cite{yu2024free}. The total elastic force exerted by the sheet on the  fluid at point $\mathbf{X}_i$ is obtained by taking the derivative of energy: $\mathbf{F}_{i} = -\partial (W  + E_b)/ \partial \mathbf{X}_i$. 

Assuming a no-slip boundary condition, the velocity of material points $\mathbf{X}_i$ of the sheet is equal to the fluid velocity: ${d\mathbf{X}_i}/{dt}=\mathbf{v}(\mathbf{X}_i)$. This equation was integrated in time using a forward-Euler method with a timestep $ \Delta t  = 6\times 10^{-5}\dot \gamma^{-1}$. The disk was discretized with 1152 triangular elements.  The computational method is detailed in \cite{yu2021coil, yu2022wrinkling}.
 
The capillary number $\Ca = \eta \dot\gamma a/G$ compares viscous and in-plane elastic stretching stresses. Smaller $\Ca$ indicates a less extensible disk. In the present work, we consider nearly-inextensible disks ($\Ca=0.01$) for which the observed change in disk area is less than $1\%$. The elasto-viscous number $\evn = {\eta\dot\gamma a^3}/{K_B}$ compares viscous and bending stresses. Small
$\evn$ implies a disk that resists out-of-plane deformations. 

\begin{figure*}[htb!]
    \centering
    \includegraphics[width=\textwidth]{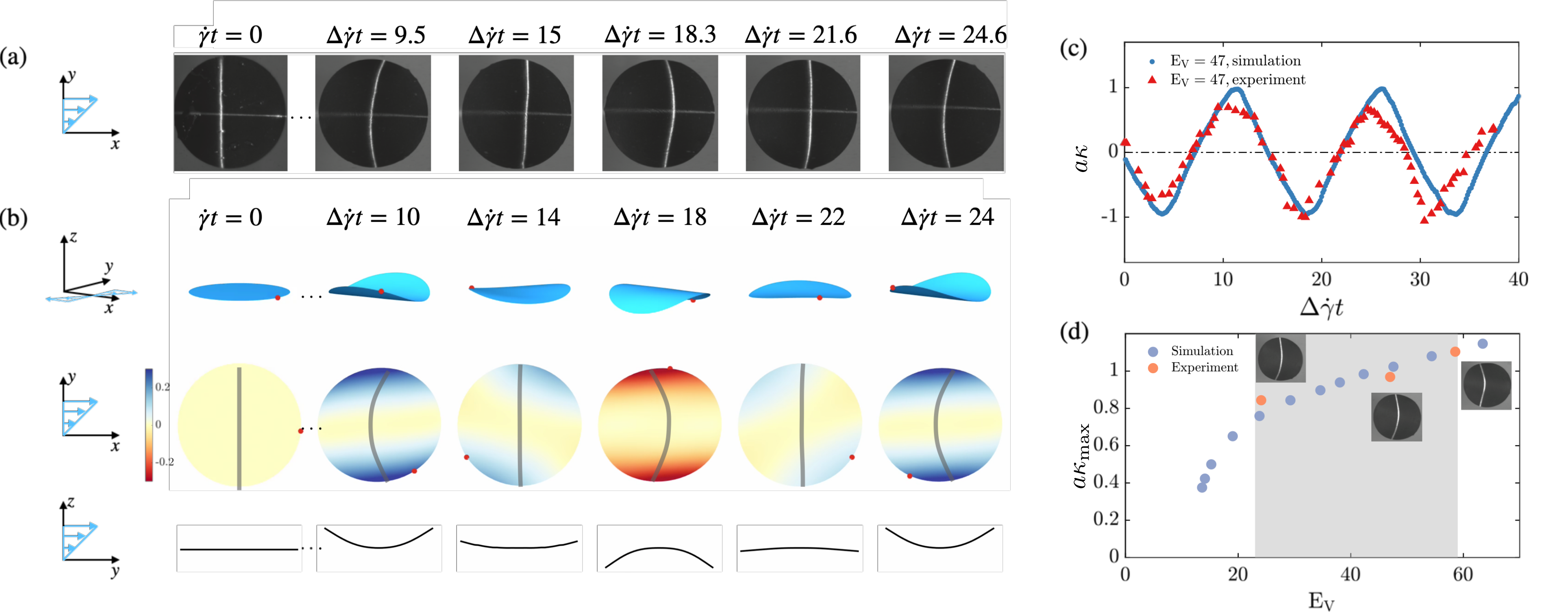}
    \caption{(a) Experimental snapshots for $\evn = 47$ illustrating the time evolution of the  flapping dynamics.  The position of the disk along $z$  and its curvature $\kappa$  are extracted from the curvature of the laser line (white in the figure).   (b) Simulation snapshots for $\evn = 47$. The perspective view illustrates the evolution of the shape of the disk. The flow-gradient ($xy$) plane view shows a heat plot of the amplitude of the out-of-plane deformation. The gray lines are the intersections of the deformed disks with inclined planes to mimic the laser sheet in the experiment. The gradient-vorticity ($yz$) plane view illustrates the sheet deformation along the line corresponding to the laser line. (c) Comparison of time evolution of curvature $\kappa$ in experiment (Fig \ref{fig:free_disk_dynamics}(a)) and simulation (Fig \ref{fig:free_disk_dynamics}(b)). The alternating sign of $\kappa$ indicates the sheet flaps up and down . (d)  The evolution of the maximum curvature $\kappa_{max}$ vs. $\evn$ enables the identification of the flapping regime. The gray region locates the range $\evn= 24 - 59$ where flapping is observed in the experiments. Simulations reveal flapping in the range $\evn= 14 - 64$. Snapshots show flapping sheets observed in experiments. }
    \label{fig:free_disk_dynamics}
\end{figure*}




{When the disk is placed in the flow with its normal vector aligned along the direction of the (undisturbed) vorticity vector, the disk rotates rigidly  when the shear rate, and thus $\evn$, is sufficiently small. However, beyond a threshold shear rate periodic deformations occur.  In Fig \ref{fig:free_disk_dynamics}(a) we show experimental images of the disk (in black) as seen from above (along the $z$ axis) for different times and $\evn\simeq47$. To characterize the time dependence of the curvature, we shine a laser sheet at a fixed angle $\simeq 45^\circ$. From the line of intersection of the surface of the disk with the laser sheet (white line in Fig \ref{fig:free_disk_dynamics}) we calculate the curvature $\kappa$ at the center of the sheet along the fixed axis $y$. The measured curvature displays oscillations around zero, with a period $ T = 15.1 \dot \gamma^{-1}$ and maximum amplitude  in absolute value $| a \kappa|   \simeq 1.0 $ (Fig \ref{fig:free_disk_dynamics}(c)).

Free-space simulations recover the flapping behaviour seen in experiments for disks with zero spontaneous curvature. To trigger the flapping instability in simulations, a slightly buckled C-shape is used as the initial condition, with an initial out-of-plane amplitude of $10^{-4}a$. Fig.~\ref{fig:free_disk_dynamics}(b) shows four perpendicular views of one cycle of a flapping motion from the simulations with $\evn = 47$. As seen in experiments, the disk curves up and down periodically. The simulations indicate that the curvature along the axis perpendicular to the line of maximum curvature is negligible, however, the surface is not exactly cylindrical. We measure a negative non-zero Gaussian curvature of the order of $0.1/a^2$. 

Material points on the surface of the disk experience hydrodynamic compression or extension depending on whether these points lie in the compressional quadrant $(x<0, y>0;x>0, y<0)$ or in the extensional quadrant $(x>0, y>0;x<0, y<0)$ of the shear flow, respectively  \cite{guyon}. By placing a red dot on the perimeter of the simulated disk (see Fig.~\ref{fig:free_disk_dynamics}(b)), we can follow the trajectory of one such material point. This trajectory indicates that portions of the disk displaying maximum out-of-plane deformation lies in the compressional quadrant, as for flexible fibers \cite{ARFLindner}. When the material axis of maximum out-of-plane deformation rotates and enters the extensional quadrant, the deformation relaxes (in blue on  Fig.~\ref{fig:free_disk_dynamics}(b), time $\dot \gamma t = 14$). In the mean time, the material axis aligned along the compressional quadrant buckles. The overall deformation mode is saddle shaped, as we measured a negative Gaussian curvature. This second out-of-plane deformation, located in the compressional quadrant and having a sign opposite to that of the out-of-plane deformation in the extensional quadrant, increases over time and becomes the new dominant direction of out-of-plane deformation. (in red on  Fig.~\ref{fig:free_disk_dynamics}(b), fourth disk). This qualitative observation suggests that the flapping instability may not be observed for an ideal inextensible disk, for which saddle shape are forbidden \cite{audoly2000elasticity}. We found that a further decrease in $\Ca$ gives no qualitative change in sheet dynamics, but produces a decrease in the flapping amplitude.

We provide a direct comparison between experiments and simulation in Fig \ref{fig:free_disk_dynamics}(c), which shows the rescaled curvature $a \kappa$ along the line of the $y-z$ plane shown in Fig \ref{fig:free_disk_dynamics}(b). 
The simulation and the experiment show good agreement in the time evolution of $ a \kappa$, and predict a similar flapping period of about $ T = 15 \dot{\gamma}^{-1}$ (Fig \ref{fig:free_disk_dynamics}(c))
Interestingly, as can be seen by following the red marker in Fig.~\ref{fig:free_disk_dynamics}, the time period of the flapping motion is smaller than the time period of the rotation of the disk.

The flapping dynamics exist for a finite range of $\evn$. In experiments, we observe flapping for $\evn$ in the range from 24 to 59. Below $\evn\simeq 24$ the disk does not deform, rotates and leave the initial shear plane. In experiments, above $\evn \simeq 59$ the disk deforms into higher buckling mode shapes and leaves the shear plane. Simulations recover the same range of values of $\evn$ than the experiments, as can be seen in Fig.~\ref{fig:free_disk_dynamics}(d) where we plot the maximum curvature of the sheet vs. $\evn$.

\begin{figure}[!h]
    \centering
    \includegraphics[width=0.5\textwidth]{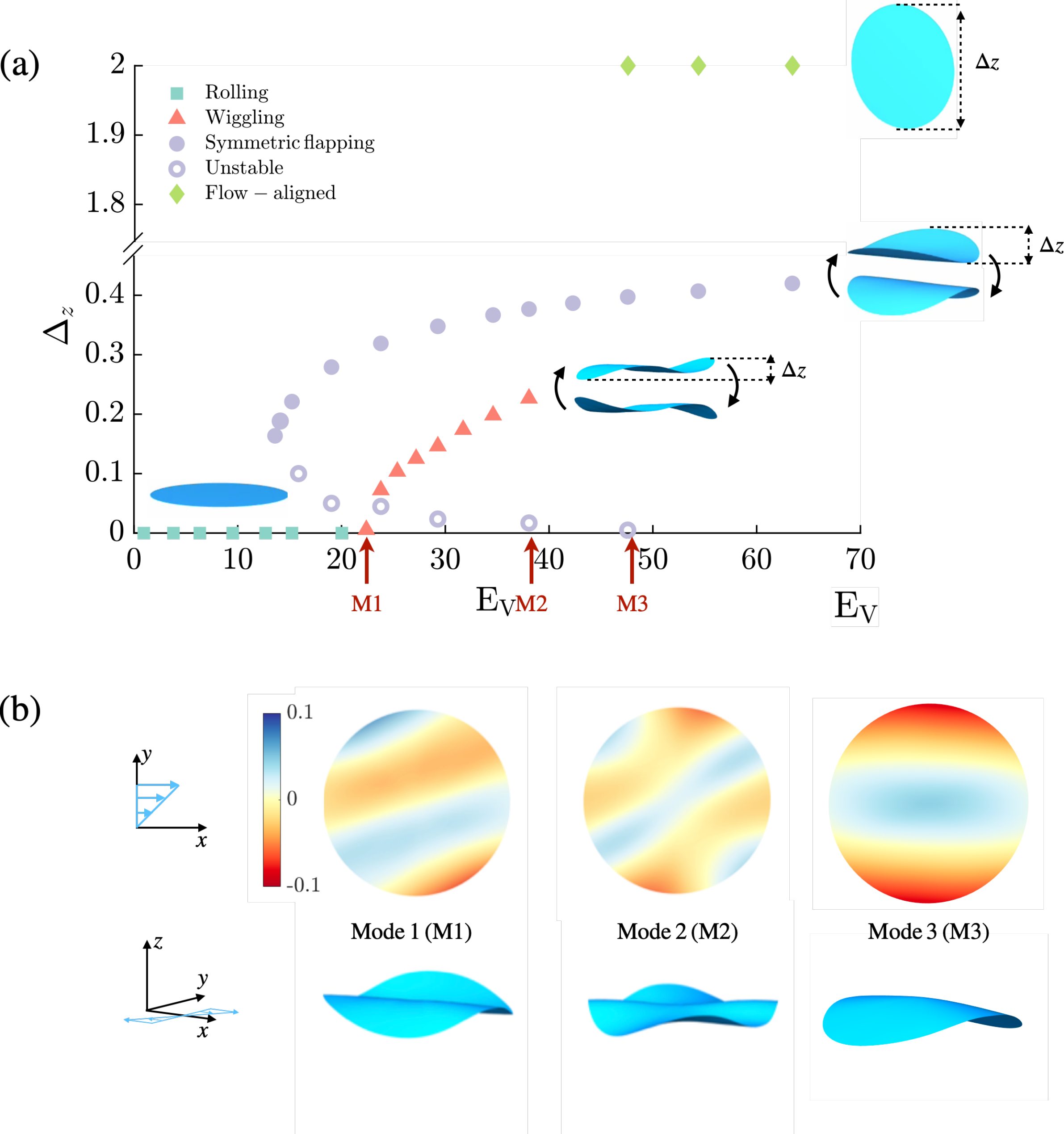}
    \caption{(a) Bifurcation diagram of deformation amplitude   
$\Delta z$ vs. $\evn$ from simulation. Snapshots describe the numerically observed dynamics within the observation window. (b) Examples of deformation modes (M1-M3) obtained from thelinear stability analysis. Flat sheet becomes linearly unstable at $\evn = 22$ for M1, $\evn = 38$ for M2, and $\evn = 48$ for M3.}
    \label{fig:free_phase}
\end{figure}

To fully investigate the dependence of the dynamics on $\evn$, on the initial condition and the perturbations, we performed time-integration of the computational model over a broad range of $\evn$ for two types of initial perturbations: (1) small random shape perturbations by randomly moving nodes with an out-of-plane magnitude of $10^{-4}a$ and (2) perturbations with a ``C-shape" similar to the observed flapping shape described above with an adjustable magnitude $\Delta_{z,0}/a$. The long-term dynamics are characterized based on a deformation amplitude $\Delta z=\max_\textrm{nodes}z-\min_\textrm{nodes} z$ measured after 800 strain units where the dynamics become steady, as shown in Fig \ref{fig:free_phase}(a).

The simulations reveal a more complex phase diagram than the experiments, and suggest the existance of a further deformation mode which we call ``wiggling'. For $\evn < 22$, small initial random shape perturbations relax to a flat shape and an in-plane rolling motion. At $\evn = 22$, the in-plane rolling motion becomes linearly unstable (as confirmed by the linear stability analysis below) and exhibits a supercritical linear buckling instability: the disk evolves into an S-shaped deformation whose maximum curvature axis oscillates by switching buckling direction dynamically as the disk rotates. With increasing $\evn$, the wiggling motion is still observed but shows a larger deformation amplitude. For  $\evn > 38$, wiggling becomes unstable and the disk reorients, assuming a flat shape with a normal in the flow plane. In this configuration, the sheet rotates until the normal to the sheet is perpendicular to the flow direction. Once it reaches this orientation, the sheet stops rotating (rotations away from this aligned state are impossible in the simulation due to the zero hydrodynamics thickness assumption, just as an infinitely thin rigid fiber in shear will permanently align with the flow direction).

When subjected to a finite-amplitude C-shaped initial perturbation, a sheet with $\evn < 14$ evolves back to a flat shape and rolling motion. For $\evn = 14$, a sufficiently large initial perturbation amplifies in time and the sheet exhibits stable periodic flapping. We identified two out-of-plane solution branches arising from $\evn = 14$,  indicated in Fig \ref{fig:free_phase}(a), corresponding to a saddle-node bifurcation. The first branch (see the gray plain circles on  fig.\ref{fig:free_phase}(a))is the stable flapping solution, whose flapping amplitude $\Delta z$ increases with $\evn$. Additionally, we found an unstable branch (see the gray empty circles on  fig.\ref{fig:free_phase}(a)) that represents the amplitude perturbation $\Delta z$ required to evolve toward stable flapping. Below the unstable branch, the initial C-shaped deformation decays in time, and the dynamics are governed by the linear buckling instability: the sheet either stays flat ($\evn < 22$) or evolves to the wiggling branch ($\evn > 22$). 

In contrast to the symmetric flapping observed in the free space simulation, the sheet in experiments shows a more pronounced flapping downward (along $-z$ axis) shown in Fig~\ref{fig:free_disk_dynamics} (d), resulting in a more negative evolution in $\kappa$. The asymmetry observed in experiments could be attributed to the presence of a free surface at a finite distance from the sheet in the experimental setup. To test this hypothesis, we performed additional simulations incorporating a nearby free surface by imposing no-penetration and no-shear boundary conditions to represent the free surface, consistent with the experimental configuration~\cite{yu2023dynamics}.  
These simulations reproduced a similar asymmetric flapping behavior. The presence of the free surface breaks the symmetry of the flow field and promotes the development of a C-shaped deformation from an initially flat sheet, thereby facilitating the evolution towards flapping.

The buckling instability of the flat state described above can be analyzed in further detail by a linear stability analysis; this also provides insight into further modes of instability and their impact on long-term behavior. Because the motions of material points in the rolling flat disk are time-periodic, this analysis makes use of Floquet theory \cite{IoossJoseph}: i.e. we investigate the response of the periodic solution  $\mathbf{X}_i(t) = \mathbf{X}_i(t+T)$ with period $T$ to infinitesimal perturbations. Linearization of the system gives eigenvalues $\sigma$ and eigenmodes that characterize the linear stability of the system. Unstable perturbation patterns with a positive growth rate $Re(\sigma) > 0$  correspond to a deviation of the flat disk shape solution of a 2D rolling disk. In Fig \ref{fig:free_phase}(b), we show the first three unstable modes. Mode M1, which emerges at $\evn \approx 22$, corresponds to a rotation symmetric S shape , consequence of a linear buckling instability. The S mode arises from the linear stability analysis agrees with the threshold $\evn$ observed in the simulations, for which the trivial 2D rolling becomes unstable, see Fig \ref{fig:free_phase}(a). Mode M2, which appears at $\evn = 38$, corresponds to an M shaped deformation. Mode M3 arises at $\evn = 48$ and corresponds to a C shape.

The linear analysis  helps explaining the observed dynamics. For instance, the time evolution of the S mode yields the wiggling solution that is observed in the simulation. In the simulation, the sheet experiences small, random initial perturbations. The parameter regime where only the S mode arises ($22 < \evn < 38$) reaches stable wiggling as it is the only unstable direction to evolve. The evolution of the M mode ($38 < \evn < 48$) in the linear regime evolves into a dynamic M-shape flapping. However, due to the interaction of this mode with the dominant S mode, the long-term evolution of the M mode into wiggling or flapping depends on $\evn$. This shows the limit of the linear study as it cannot precisely capture the nonlinear dynamics with multiple unstable modes. When the third unstable mode arises ($\evn > 48$), the nonlinear interaction between unstable modes destabilizes periodic deformations and triggers the sheet to become flow-aligned.

In conclusion, we demonstrated both numerically and experimentally that for an elastic thin disk initially aligned parallel to the shear (flow-gradient) plane, a flapping dynamics develops whereby the sheet shows, for a finite range of flow strengths  $\evn$, periodic C-shaped deformations. The flapping is a result of periodic compressive and tensile stresses due to the imposed shear flow that combine in a non-trivial manner with the finite extensibility of the sheet, which allows a non-zero Gaussian curvature.
The simulations enabled us to establish a full bifurcation diagram. The stable flapping branch is demonstrated to be part of a saddle-node bifurcation that also includes a supercritical linear instability leading to a wiggling motion. Only for small values of $\evn$ the disk rolls trivially as expected from the geometric symmetry of the disk. For large values of  $\evn$ the particle can change alignment. 

More generally, our investigation of the elasto-viscous deformation dynamics and long-time orientational behavior of a disk contributes to the understanding of shear-induced morphologies of thin, sheet-like soft particles in flow. It can serve as a stepping stone for investigating more complex systems, such as suspensions of flexible 2D crystalline materials, 2D polymers, or sheet-like macromolecules of biological origin.



\begin{acknowledgments}
This material is based on work supported by the National Science Foundation under grant No.~CBET-1604767. This project has received funding from the European Research Council (ERC) under the European Union’s Horizon 2020 research and innovation program (Grant agreement No. 715475)
\end{acknowledgments}


\bibliography{shear.bib}

\appendix

\section{Supplemental Materials}
\section{Rheology of the silicon elastomer}
We present the rheological curves of the silicon elastomer filled with carbon black, measured with an Anton Paar MCR 501 rheometer, see figure \ref{fig:Gblack}.  The asymptotic value of the storage modulus at vanishing pulsation is $G' \simeq 123 \;\rm{kPa}$. The Young modulus is $E~=~2(1~+~ \nu)G'~\simeq~370\;\rm{kPa}$ with $\nu \simeq 0.5$.
\begin{figure}[h]
    \centering
    \includegraphics[width=0.4\textwidth]{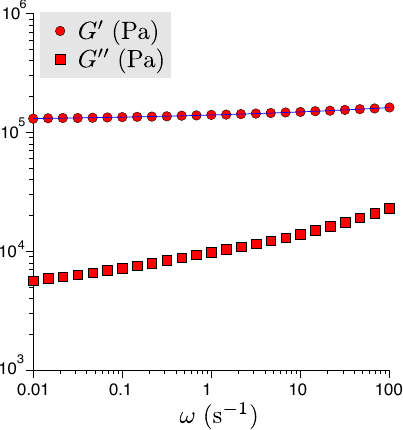}
    \caption{Rheology of the silicon elastomer filled with carbon black. Storage modulus $G'$ and loss modulus $G''$ as a function of the pulsation $\omega$.}
    \label{fig:Gblack}
\end{figure}
\subsection{Model validation: buckling dynamics of a tumbling rectangle}
We provide a validation case of the numerical model by comparing the dynamics with previous experiments for an initial condition where a flat rectangle sheet is initially placed near the flow-vorticity plane with the normal vector of the sheet inside the flow-gradient plane (see Fig \ref{fig:free_tumbling_sum}). The long edge aligns with the flow-gradient plane and is initially oriented $10^\circ$ relative to the flow direction.
\begin{figure}[h]
    \centering
    \includegraphics[width=0.5\textwidth]{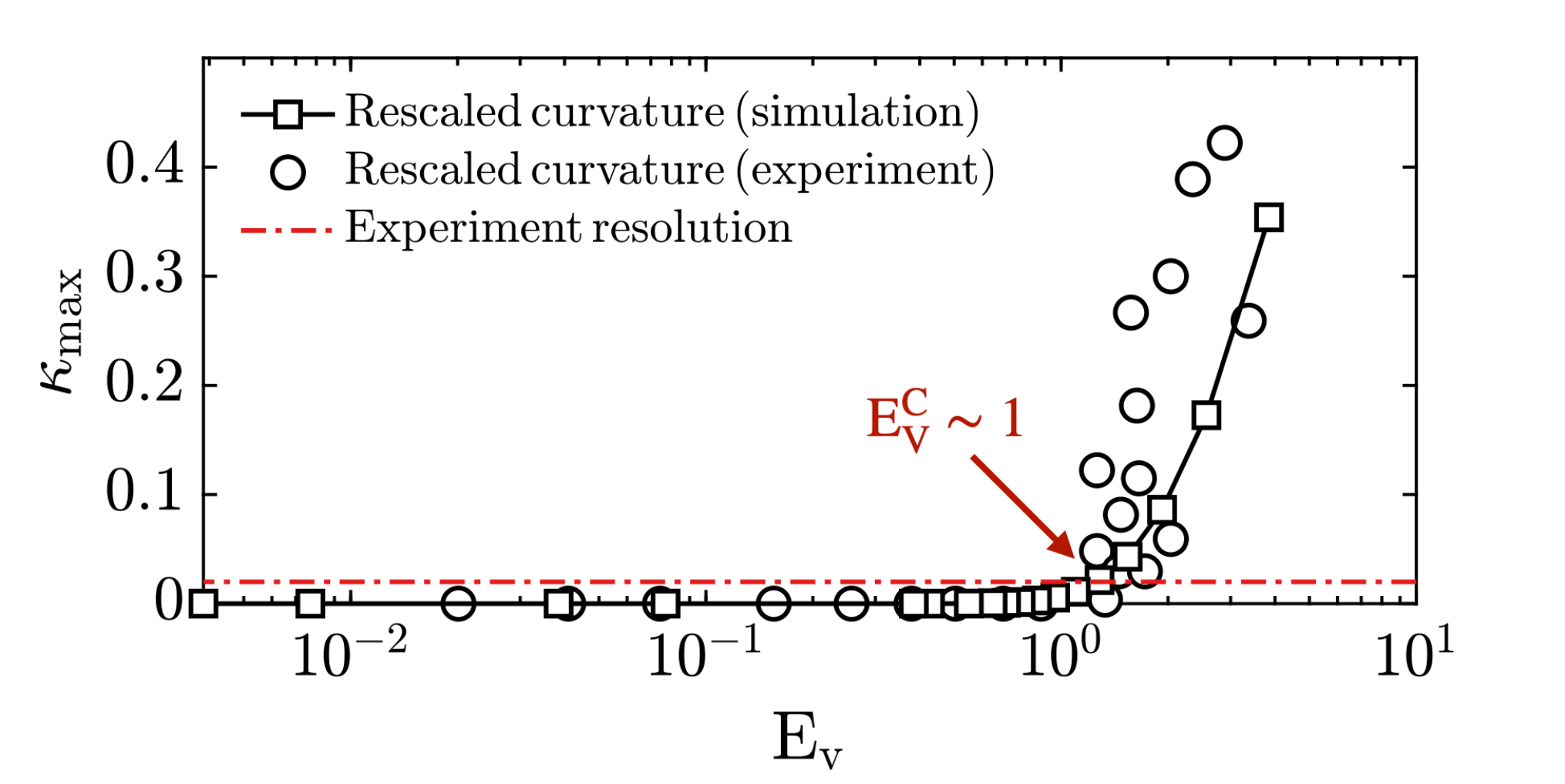}
    \caption{Maximum curvature during tumbling $\kappa_{\mathrm{max}}$ vs. $\evn$ in experiments with curvature measurement resolution $\kappa \sim 0.02$. The buckling threshold is determined as $\evnc \sim 1$.}
    \label{fig:free_tumbling_sum}
\end{figure}
Here, we compare the bucking threshold $\evnc$ where the sheets buckle due to compressive stresses in shear flow. The threshold is determined based on the experimental resolution for curvature measurement (see Fig \ref{fig:free_tumbling_sum}). The results indicate a good agreement between experiments and simulation, with $\evnc \sim 1$.

 \subsection{Linear stability analysis}


The linear stability analysis focuses on finding unstable perturbations that make dynamics deviate from the base solution: a disk undergoing 2D rolling in the shear plane. We describe the conformation of the disk by stacking all the material point nodal positions $\mathbf{X}_i$ into a vector $\mathbf{X}$. For the rolling solution, the material points on the disk follows periodic trajectories with period $T$, so
\begin{equation}
	\mathbf{X}(t) = \mathbf{X}(t+T).
\end{equation}
We use Floquet theory to examine the linear stability of of this motion \cite{IoossJoseph}. Along the trajectory, we arbitrarily choose an origin for time, and apply a random initial deformation pattern $\hat{\mathbf{X}}$ with small magnitude $O(\varepsilon) \sim 10^{-4}$:
\begin{equation}
	\mathbf{X} = \mathbf{X}(t) +\varepsilon\hat{\mathbf{X}}.
\end{equation}
Because $\varepsilon$ is very small, the system evolution remains linear over one time period, as the deformation in that time is small. Therefore, we treat the time integration over one period as a linear operator $\mathbf{A} = e^{\mathbf{L}T}$ acting on the $\mathbf{X}(t)$:
\begin{equation}
	\hat{\mathbf{X}}(t+T) = \mathbf{A}\hat{\mathbf{X}}(t) = e^{\mathbf{L}T}\hat{\mathbf{X}}(t).
\end{equation}
We then extract eigenvalues $\sigma$ and modes of $\mathbf{L}$ repetitively via a modified Arnoldi iteration method~\cite{trefethen2022numerical,yu2022wrinkling}. A stable perturbation ($Re(\sigma) < 0$) decays exponentially after completing one period, while an unstable perturbation ($Re(\sigma) > 0$) grows.


\end{document}